\documentstyle[mncite,draft,epsf]{mn}

\newcommand{\vsini}{\mbox{$v_e\,\sin\,i$}}

\newcommand{\lii}{Li\,{\footnotesize I}}
\newcommand{\fei}{Fe\,{\footnotesize I}}

\newcommand{\ki}{K\,{\footnotesize I}}
\newcommand{\cai}{Ca\,{\footnotesize I}}
\newcommand{\kms}{\,km\,s$^{-1}$}

\newcommand{\be}{\begin{equation}}
\newcommand{\ee}{\end{equation}}
\newcommand{\bd}{\begin{displaymath}}
\newcommand{\ed}{\end{displaymath}}

\title[Lithium in the Pleiades]{On the lithium abundance dispersion in
late-type Pleiades stars}

\author[R. D. Jeffries] {R. D. Jeffries \\
Department of Physics, Keele University, Keele, Staffordshire,
ST5 5BG, UK\\
}
\date{Received 18 March 1998}
\pagerange{\pageref{firstpage}--\pageref{lastpage}}
\def\LaTeX{L\kern-.36em\raise.3ex\hbox{a}\kern-.15em
    T\kern-.1667em\lower.7ex\hbox{E}\kern-.125emX}

\begin{document}

\label{firstpage}

\maketitle

\begin{abstract}
I present the results of a programme to monitor the strengths of the
\lii\ 6708\AA, \ki~7699\AA\ and chromospheric H$\alpha$ lines in a
group of cool Pleiades stars. Consistent instrumentation and analysis
techniques are used to show that there is no \lii\ variability on
timescales of 1 year that could possibly account for the apparent
spread in Li abundances seen in Pleiades stars between effective
temperatures of 4800-5200\,K.  Comparison with published data reveals
tentative evidence for variability on 10 year timescales, but at a very
low level. The lack of chromospheric activity variability above levels
of 20 to 30 percent makes it difficult however, to rule out evenly
distributed magnetic activity regions causing a scatter in the \lii\
line strengths at a given abundance.
The similar star to star scatter of \ki\ line strengths in
these and published data reinforces the conclusion that it is still
unsafe to attribute the \lii\ line strength dispersion to a large
variation in Li depletion at a given mass.
\end{abstract}


\section{Introduction}

Lithium is unique among metals -- it is produced in the big bang, but
is destroyed by p,$\alpha$ reactions at low temperatures ($\sim
2.4\times10^{6}$\,K) in stellar interiors.  Observing Li in the cool
stars of young open clusters is an excellent way of exploring and
calibrating the physics of internal stellar mixing due to convection or
additional ``non-standard'' processes. Rapid angular momentum loss as
stars reach the ZAMS may make Li depletion due to poorly understood
mechanisms such as rotational mixing, much more dramatic.  Such
investigations are important in understanding stellar physics, but also
in deciding whether present day Population II Li abundances have hardly
changed from the primordial big bang nucleosynthesis value. This is
supported by ``standard'' models that feature {\em only} convective
mixing (Deliyannis, Demarque \& Kawaler 1990, Bonifacio \& Molaro
1997), but other models including ``non-standard mixing'' predict that
the Population II Li has been depleted from primordial values a factor
of 2-3 higher (Deliyannis \& Ryan 1997, Pinsonneault et al. 1998).

A particular problem for the standard
models that has emerged in the last 15 years is the observation that
cool stars (4500\,K$<T_{eff}<$5400\,K) in the Pleiades (age 100\,Myr)
and Alpha Per (age 50 Myr) clusters exhibit an apparent Li abundance
spread of more than an order of magnitude at a given mass (Duncan \&
Jones 1983, Butler et al. 1987, Soderblom et al. 1993a [hereafter - S93a], 
Balachandran, Lambert \& Stauffer 1996,
Randich et al. 1998). This spread cannot be
reconciled with standard models, which predict a unique amount of
pre-main sequence (PMS) Li depletion at a given mass and age. 

Several explanations have been put forward which assume that the observed
spread corresponds to a {\em real} dispersion in the Li
abundances. These have been guided by the correlation (albeit an
imperfect one) of reduced Li depletion and rapid rotation (S93a,
Mart\'{i}n 1997, Randich et al. 1998).  Extra mixing as stars spin down
towards the ZAMS is one possibility (Chaboyer et al. 1995). The
observed spread in rotation rates at the ZAMS could lead
to different efficiencies for rotationally induced mixing. Unfortunately,
even the most recent models, which incorporate early angular momentum
loss via a circumstellar disk, seem unable to explain the size of the
abundance dispersion (Pinsonneault et al. 1998). 
Others have postulated that structural changes associated with rapid
rotation (Mart\'{i}n \& Claret 1996) or dynamo induced magnetic fields
in the convection zone (Ventura et al. 1998) might reduce the
efficiency of standard PMS Li depletion in rapid rotators. Other
combinations of mechanisms have been plausibly suggested, such as
mixing by internal gravity waves plus the magnetic blocking 
of mechanical flux through
the bottom of the convective zone in rapid rotators
(Schatzman 1993, Montalb\'{a}n \& Schatzman 1996).
These ideas seem promising because the observed Li depletion is
significantly less than the conservative predictions due to standard
PMS convective mixing alone (Jeffries \& James 1999).
 
Another school of thought is that the {\em apparent} Li abundance
dispersion might largely be explained in terms of a scatter in the
equivalent width (EW) of the \lii\ 6708\AA\ line (almost exclusively
used for this work) at a given Li abundance -- caused by NLTE,
chromospheres, or large-scale photospheric inhomogeneities such as plages
or starspots. It has long been known that the reduced temperature in
sunspots increases the observed EW of the \lii\ 6708\AA\ line (Giampapa
1984). Young, cool stars are known to have much larger photospheric
inhomogeneities than the Sun, which are associated with their enormous
magnetic activity. The most recent modelling work (Carlsson et
al. 1994, Houdebine \& Doyle 1995, Stuik, Bruls \& Rutten 1997)
indicates that these effects may indeed induce a scatter in the
apparent Li abundances if, as is normally the case, homogeneous, one
dimensional atmospheres and $T_{\rm eff}$-colour relationships are used
to convert observed EWs into abundances. Impetus has been given to these
explanations by two pieces of observational work. S93a
(see also Stuik et al. 1997) find that the \ki\ 7699\AA\
line, formed in similar conditions to the \lii\ line, also shows a
rotation-dependent EW scatter (albeit smaller than that exhibited by \lii)
at a given colour. As no potassium abundance variations are expected in
the Pleiades, this result challenges our understanding of how neutral
alkali lines are formed in cool stellar atmospheres. Russell (1996)
examined the subordinate line of \lii\ at 6104\AA\ in several late-type
Pleiads. He argued that Li abundances from this line are lower and show
considerably less scatter than from the \lii\ 6708\AA\ line, possibly
because of different sensitivities to NLTE conditions and chromospheric
activity. Russell's conclusions have since been challenged because of the
difficulty in deblending the weak 6104\AA\ line from stronger nearby
lines (Mart\'{i}n 1997).

A useful test of whether Li abundance variations are real
is to search for variability in the \lii\ 6708\AA\ EW. If the scatter
in Li abundances is caused by large-scale atmospheric inhomogeneities one might
expect to see EW variations associated either with magnetic activity
cycles or with the rotational modulation of spots and plages. Spots are
often demonstrably asymmetric in their surface distribution 
on young, active stars, which for instance, allows rotation periods to
be measured from light curve modulations (O'Dell \& Collier-Cameron
1993). The search for Li EW variability has a chequered history and is
reviewed by Fekel (1996). There are individual examples of young active
field stars where EW variations on timescales of the rotation period have
been reported ({\em e.g.} Robinson, Thompson \& Innis 1986, Basri,
Mart\'{i}n \& Bertout 1991, Patterer et al. 1993, Jeffries et al. 1994,
Eibe et al. 1999) and other studies which claim to detect no variations
at the 5\% level or less ({\em e.g.} Boesgaard 1991, Pallavicini et
al. 1993). The only study carried out so far in young clusters is that
of S93a. They say that observations of
some cool Pleiades stars taken on several occasions show no EW
variations beyond their estimated errors of $\sim$20\,m\AA,
but no details are given. As the \lii\ 6708\AA\ EWs in late-G and
early-K type Pleiads range from $\sim$100\,m\AA\ to $\sim$300\,m\AA\,
this amount of variability would seem unable 
to account for any significant fraction of the apparent Li abundance
dispersion.  

To put this important result on a firmer footing, I have begun a
programme to monitor the \lii\ 6708\AA\ line in a selection of late-G
and early-K type Pleiads over long timescales; to test for variability,
to see whether any variability is also present in the \ki\ 7699\AA\ line
and to see whether any variability is correlated with changes in
chromospheric activity. In this short paper I present the first results of
this programme which allow a precise investigation of variability on 1 year
timescales using data taken with identical instrumentation and on
$\sim$10 year timescales for the first time, by comparison with the
results of S93a.

\section{Observations and Analysis}

\begin{center}
\begin{table*}
\caption{A log of the INT spectroscopic observations and EW
measurements.}
\begin{tabular}{lrrrrrrrrrrr}
\hline
&&&&&&&&&&\\
Name & $(B-V)_{0}$ & \vsini & Date & UT & \multicolumn{4}{c}{$\lambda\lambda 6510-6755$} &
\multicolumn{2}{c}{$\lambda\lambda 7570-7790$} \\
     &      &&&    & S/N  & \lii 6708\AA & \cai 6718\AA & Excess H$\alpha$ & S/N
& \ki 7699\AA \\
     &      & \kms&   &     & & EW (m\AA)     & EW (m\AA)     & EW (m\AA)       &                       
& EW (m\AA)      \\
&&&&&&&&&&\\
\hline
&&&&&&&&&&\\
Hz 34   & 0.89 & $<7$ & 10 Nov 1997 & 20:30 & 90 & $154\pm7$ & $186\pm6$ & $158\pm12$
& 60 & $261\pm13$ \\
Hz 345  & 0.81 & 18& 10 Nov 1997 & 21:00 & 80 & $282\pm10$& $176\pm7$ &
$1508\pm15$ & 60 & $336\pm12$ \\
Hz 174  & 0.81 & 28 & 10 Nov 1997 & 21:30 & 90 & $299\pm12$& $220\pm14$ &
$1770\pm30$ & 60 & $391\pm24$ \\
Hz 916  & 0.83 & $<7$ & 10 Nov 1997 & 22:15 & 95 & $187\pm9$& $180\pm7$
& $217\pm15$ & 60 & $282\pm11$ \\
Hz 2407 & 0.91 & $<7$ & 10 Nov 1997 & 23:00 & 100 & $116\pm7$ &
$218\pm6$ & $340\pm9$ & 65 & $317\pm12$ \\
Hz 2034 & 0.93 & 75 & 11 Nov 1997 & 00:00 & 85 & $220\pm13$ &
$238\pm14$ & $1060\pm50$ & 70 & $469\pm16$ \\ 
&&&&&&&&&&\\
Hz 345  & 0.81 & 18   & 26 Nov 1998 & 20:30 & 100 & $270\pm5$ &
$172\pm6$ & $1250\pm15$ & 60 & $337\pm15$ \\
Hz 174  & 0.81 & 28   & 26 Nov 1998 & 21:00 & 90 & $290\pm11$ &
$203\pm14$ & $1750\pm50$ & 55 & $433\pm32$ \\
Hz 916  & 0.83 & $<7$ & 26 Nov 1998 & 22:00 & 110 & $185\pm4$ &
$169\pm6$ & $207\pm12$ & 60 & $261\pm15$ \\ 
Hz 2407 & 0.91 & $<7$ & 26 Nov 1998 & 23:00 & 95 & $125\pm5$ &
$210\pm8$ & $351\pm9$ & 55 & $324\pm16$ \\
Hz 2034 & 0.93 & 75   & 27 Nov 1998 & 00:00 & 80 & $230\pm19$ &
$253\pm13$ & $1490\pm80$ & 60 & $452\pm20$ \\ 
Hz 34   & 0.89 & $<7$ & 27 Nov 1998 & 01:30 & 110 & $146\pm5$ &
$191\pm5$ & $169\pm10$ & 60 & $277\pm10$ \\
Hz 263  & 0.84 & 10   & 27 Nov 1998 & 02:40 & 100 & $240\pm5$
&$166\pm6$ & $359\pm 11$ & 65 & $300\pm10$ \\
Hz 1095 & 0.86 & $<7$ & 27 Nov 1998 & 03:30 & 90 & $130\pm6$ &
$187\pm9$ & $180\pm 13$& 55 & $285\pm15$ \\
Hz 1124 & 0.91 & 7.5  & 27 Nov 1998 & 04:30 & 90 & $189\pm7$ &
$253\pm9$ & $527\pm 15$ & 60 & $335\pm12$ \\
&&&&&&&&&&\\
\hline
\end{tabular}
\end{table*}
\end{center}

\subsection{Target selection}

A dozen Pleiades stars were selected for re-observation from Table~1 in
S93a. They were chosen to have $4800<$T$_{\rm
eff}<5200$\,K, based upon the colour-T$_{\rm eff}$ relation used by S93a
in their analysis, were not known spectroscopic binaries and had a large
range of projected equatorial velocities (\vsini) and \lii\ 6708\AA\
EWs.

\subsection{Spectroscopy}

The spectroscopy was performed at the 2.5-m Isaac Newton Telescope
(INT) on the nights of 1997 November 10 and 1998 November 26. In both
cases precisely the same instrumental set up was used. The
Intermediate Dispersion Spectrograph, 500\,mm camera, H1800V grating
and a TEK 1024 pixel square detector were used to cover the spectral
region from 6510\AA\ to 6755\AA\ at a dispersion of 0.23\AA\ per
24\,$\mu$m pixel. A 1 arcsec slit projected to just over 2 pixels on
the CCD, giving an instrumental spectral resolution (confirmed by arc
lines) of 0.48\AA. A second wavelength setting was also used that
covered the range 7570\AA\ to 7790\AA\ at 0.22\AA\ per pixel and an
instrumental resolution of 0.46\AA.

The usual calibration bias frames, tungsten flat fields and B star spectra
were taken, as well as copper/neon lamp exposures at every target
position.  A set of ``minimum chromospheric activity'' standard stars, from the
list in Soderblom et al. (1993b), were also observed at high
signal to noise (S/N).
The data were reduced and spectra extracted and wavelength
calibrated using the Starlink {\sc figaro} software package. Two
exposures of each Pleiades target were taken at each wavelength setting. The
exposure times were judged to yield S/N levels of
80-110 per pixel in the continuum of the co-added \lii\ 6708\AA\
spectra and 55-70 per pixel in the continuum of the co-added \ki\
7699\AA\ spectra.
 
The seeing was not particularly good on either night (1.5-2.5 arcsec)
and the final 4 hours of the 1997 observations were badly cloud affected.
Nevertheless I was able to complete observations of six of the
programme targets in 1997 and nine in 1998. A log of the observations
and the S/N levels achieved in each co-added spectrum is given in
Table~1.

EWs of the \lii\ 6708\AA, \cai\ 6718\AA\ and \ki\ 7699\AA\ lines were
measured by direct integration below continuum levels which were
determined from polynomial fits to approximately line free regions of the
spectra. For targets where I have a pair of observations from 1997 and
1998, I was careful to use {\em identical} continuum regions, after
correcting the spectra to a similar rest velocity, so that
EW measurements are directly comparable for these stars. Note that it
was not possible to select identical continuum regions for all the
targets because the rapid rotators suffer from extensive blending and
consequent blanketing of the continuum. The \ki\ line is affected by a
small amount of telluric absorption. This was corrected for by reference
to a B-star spectrum.

I also measured the H$\alpha$ feature as an estimator of chromospheric
activity. This is usually in
absorption in inactive stars, but can be filled in or even show
emission in the most active G/K stars. Again, consistent continuum
measurements were used in pairs of observations to aid comparison. 
Following Soderblom et al. (1993b), I subtracted the spectrum of each
Pleiades star from the spectrum of a normalised minimum activity star
of similar intrinsic colour. The standards used are the same as those
in Soderblom et al. (1993b), namely, HD 3651, HD 4256, HD 4628, HD
10476, HD 166620 and HD 182488. For the two stars with \vsini$>20$\kms,
the standard star was broadened to the appropriate \vsini, before
subtraction. The excess EW of the resultant ``emission line'' can be taken as
a measure of the chromospheric activity and is comparable with
the measurements in Soderblom et al. (1993b). Again, telluric features
were visible in some of the spectra and were removed by reference to a scaled
high S/N spectrum of a rapidly rotating B-star.

\begin{table}
\caption{A comparison of deblended \lii\ 6708\AA\ EWs from this paper and
Soderblom et al. (1993a).}
\begin{tabular}{lrrr}
\hline
&&&\\

Name & \multicolumn{3}{c}{Deblended \lii\ EW (m\AA)}\\
     & S93a & Nov 1997 & Nov 1998\\
&&&\\
\hline
Hz 34  &$134\pm 18$&$139\pm 7$ &$131\pm 5$ \\
Hz 174 &$241\pm 25$&$286\pm 12$&$277\pm 11$\\
Hz 263 &$290\pm 12$&--         &$226\pm 5 $\\
Hz 345 &$245\pm 12$&$269\pm 10$&$257\pm 5$ \\
Hz 916 &$199\pm 12$&$173\pm 9$ &$171\pm 4$ \\
Hz 1095&$138\pm 12$&--         &$116\pm 6$ \\
Hz 1124&$217\pm 18$&--         &$174\pm 7$ \\
Hz 2034&$222\pm 36$&$204\pm 13$&$214\pm 19$\\
Hz 2407&$125\pm 12$&$101\pm 7$ &$110\pm 5$ \\
\hline
\end{tabular}
\end{table}

\begin{table}
\caption{A comparison of \cai\ 6718\AA\ EWs from this paper and
Soderblom et al. (1993a).}
\begin{tabular}{lrrr}
\hline
&&&\\

Name & \multicolumn{3}{c}{\cai\ EW (m\AA)}\\
     & S93a & Nov 1997 & Nov 1998\\
&&&\\
\hline
Hz 34  &$186\pm 18$&$186\pm 8$ &$191\pm 5$ \\
Hz 174 &$269\pm 25$&$220\pm 14$&$203\pm 14$\\
Hz 263 &$218\pm 12$&--         &$166\pm 6 $\\
Hz 345 &$182\pm 12$&$176\pm 7$&$172\pm 6$ \\
Hz 916 &$203\pm 12$&$180\pm 7$ &$169\pm 6$ \\
Hz 1095&$216\pm 12$&--         &$187\pm 9$ \\
Hz 1124&$237\pm 18$&--         &$253\pm 9$ \\
Hz 2034&$256\pm 38$&$238\pm 14$&$253\pm 13$\\
Hz 2407&$229\pm 12$&$218\pm 6$ &$210\pm 8$ \\
\hline
\end{tabular}
\end{table}

\begin{table}
\caption{A comparison of \ki\ 7699\AA\ EWs from this paper and
Soderblom et al. (1993a).}
\begin{tabular}{lrrr}
\hline
&&&\\

Name & \multicolumn{3}{c}{\ki\ EW (m\AA)}\\
     & S93a & Nov 1997 & Nov 1998\\
&&&\\
\hline
Hz 34  &$233\pm 18$&$261\pm 13$ &$277\pm 10$ \\
Hz 174 &--         &$391\pm 24$&$433\pm 32$\\
Hz 263 &$250\pm 12$&--         &$300\pm 10 $\\
Hz 345 &$321\pm 12$&$336\pm 12$&$337\pm 15$ \\
Hz 916 &$222\pm 12$&$282\pm 11$ &$261\pm 15$ \\
Hz 1095&$235\pm 12$&--         &$285\pm 15$ \\
Hz 1124&$369\pm 18$&--         &$335\pm 12$ \\
Hz 2034&--         &$469\pm 16$&$452\pm 20$\\
Hz 2407&$294\pm 12$&$317\pm 12$ &$324\pm 16$ \\
\hline
\end{tabular}
\end{table}

\begin{table}
\caption{A comparison of {\em excess} H$\alpha$ EWs from this paper and
Soderblom et al. (1993b). Where several measurements were given by
Soderblom et al. (1993b), all are listed.}
\begin{tabular}{lrrr}
\hline
&&&\\

Name & \multicolumn{3}{c}{Excess H$\alpha$ EW (m\AA)}\\
     & S93b & Nov 1997 & Nov 1998\\
&&&\\
\hline
Hz 34  &$230\pm 15$ &$158\pm 12$ &$169\pm 10$ \\
Hz 174 &--          &$1770\pm 30$&$1750\pm 50$\\
Hz 263 &$480\pm 15$ &--          &$359\pm 11 $\\
       &$340\pm 15$ &&\\
       &$390\pm 15$ &&\\
Hz 345 &$1570\pm 15$&$1508\pm 15$&$1250\pm 15$ \\
       &$1050\pm 15$&&\\
       &$1240\pm 15$&&\\
Hz 916 &$280\pm 15$ &$217\pm 15$ &$207\pm 12$ \\
       &$290\pm 15$ &&\\
Hz 1095&$140\pm 15$ &--          &$180\pm 13$ \\
       &$340\pm 15$ &&\\
Hz 1124&$580\pm 15$ &--          &$527\pm 15$ \\
Hz 2034&--          &$1060\pm 50$&$1490\pm 80$\\
Hz 2407&$230\pm 15$ &$340\pm  9$ &$351\pm 9$ \\
\hline
\end{tabular}
\end{table}

\section{Results}

\begin{figure}
\vspace*{9cm} \includegraphics{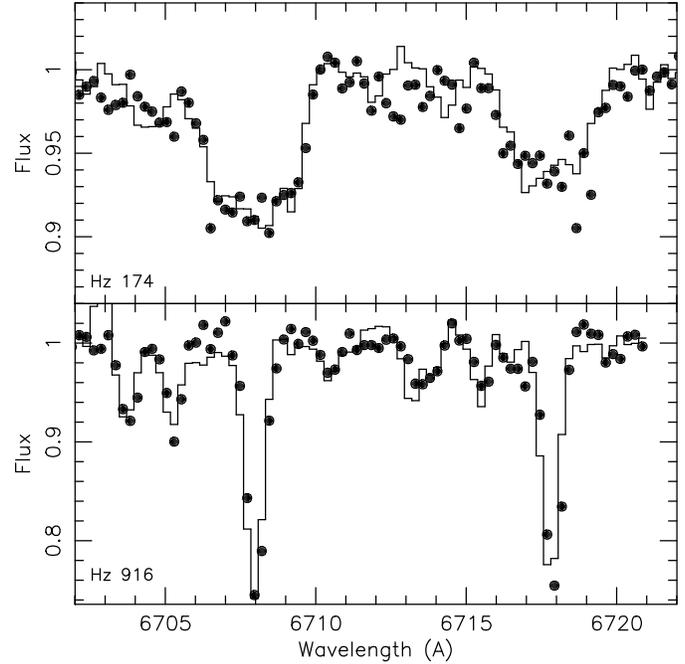}
\caption{Spectra for Hz 174 and Hz 916 
around the \lii\ 6708\AA\ line from 1997 (dots) and
1998 (solid lines).}
\label{fig1}
\end{figure}
\begin{figure}
\vspace*{9cm} \includegraphics{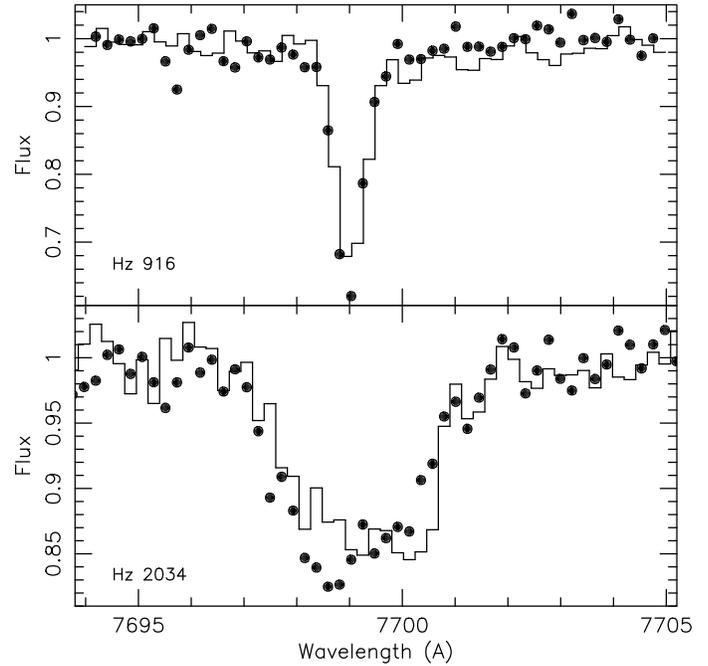}
\caption{Spectra for Hz 916 and Hz 2034 around the \ki\ 7699\AA\ line 
from 1997 (dots) and
1998 (solid lines).}
\label{fig2}
\end{figure}
\begin{figure}
\vspace*{9cm} \includegraphics{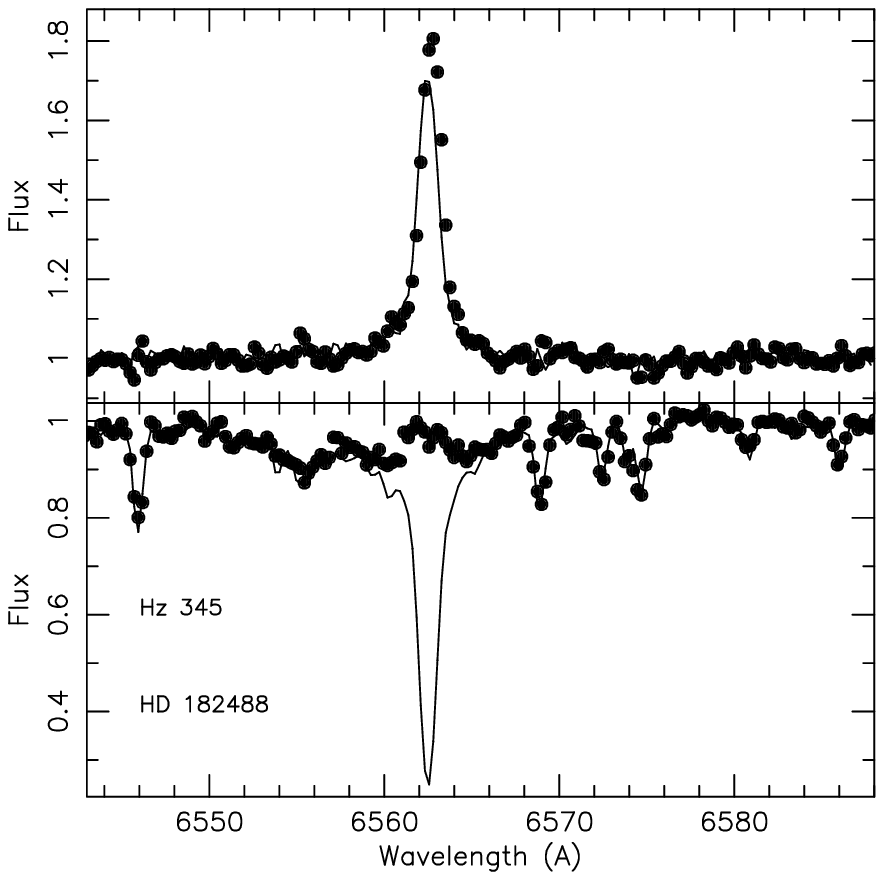}
\caption{The bottom panel shows the spectral subtraction technique
applied to Hz 345 (dots) around the H$\alpha$ line in 1998. The
template star (solid line) is HD 182488. The top panel shows the 
subtracted ``excess''
H$\alpha$ emission lines for Hz 345 in 1997 (dots) and 1998 (solid line).}
\label{fig3}
\end{figure}

The EW measurements for all the relevant lines are given in Table 1,
along with the S/N in the co-added spectra.  Tables 2-5 list these results
again and compare them with the results of S93a and Soderblom et
al. (1993b). In Table~2 I compare the ``deblended'' \lii\ 6708\AA\
EWs quoted by S93a. The blend in question is with a weak \fei\ feature at 6707.44\AA,
which has a strength of 20$(B-V)_{0}-3$\,m\AA\ according to S93a. The
EWs listed in Table 2 have all been corrected
according to this formula. Figures 1 and 2 show examples of the spectra
from 1997 and 1998 for comparison of the \lii, \cai\ and \ki\ lines. Figure 3
illustrates the spectral subtraction technique used for H$\alpha$ and
shows an example of the comparison between excess H$\alpha$
emission in 1997 and 1998.

\subsection{Variability on one year timescales}

The strength of the small dataset presented here is that the observations
were taken with identical instrumentation and reduced in a very
consistent manner. Comparison of the six Pleiades stars in common to
the November 1997 and November 1998 datasets allows a strong statement
on the amount of variability on yearly timescales, independent of any
possible systematic errors in the EW measurements. It is clear from Tables
2-4 that the EWs of the \lii\, \cai\ and \ki\ lines are completely in
agreement. For \lii\ the biggest (and most significant) 
discrepancy is ($12\pm 11$)\,m\AA\ for Hz 345. The weighted
mean discrepancy for all six stars is $(2\pm 4)$\,m\AA, with an rms discrepancy of less
than 14\,m\AA\ at a 95 percent confidence level.

The EW discrepancies can be
somewhat larger for the other two lines, although their errors are larger
and the significance of the discrepancy is still less than 1.5$\sigma$
in all cases. From the excess H$\alpha$ EWs,
there is clear evidence for variability in the chromospheric emission
of Hz~345 and Hz~2034 at the 20-30 percent level, whereas the other four
stars are constant to better than 10 percent.
 
\subsection{Variability on ten year timescales}

The dataset provided by S93a allows variability on longer timescales to
be probed. Here we have a slightly larger dataset (nine stars), but there
could be systematic differences in the EW measurements due to choice of 
continuum placement and slightly different spectral resolutions. 
Ignoring these possible systematic errors for the present, we can see
that the 1997/98 \lii\ line measurements are remarkably consistent with
those of S93a taken in 1988 and 1990. The maximum and most significant 
discrepancy is ($64\pm 13$)\,m\AA\ for Hz 263. The weighted mean discrepancy
between S93a and our 1998 observations is $(21\pm 5)$\,m\AA, in the sense that
the S93a EWs are larger. However, from the error estimates,
this cannot be explained (is statistically unacceptable) 
as a {\em systematic} continuum placement problem
and there may be genuine variation in the \lii\ EWs at an rms level of about 30\,m\AA.

A useful check on any systematic error is provided
by the nearby \cai\ line, which will have been measured with respect to
a similar adjacent continuum region. The \cai\ line is formed from an
excited level and has been shown to be considerably less sensitive to
photospheric inhomogeneities than the \lii\ line. If both the
\lii\ and \cai\ line show EW discrepancies in the same sense and of similar
size then the continuum placement may be to blame for most of the
apparent variation. From Tables 2 and 3, we can see that this may be
the case for Hz 263, Hz 916, Hz 1095 and Hz 2407, but Hz 174 and Hz
1124 show \cai\ EW discrepancies in the opposite sense to those in
\lii. If all the statistical error estimates are correct, then we have
to conclude that there is some genuine \lii\ EW variation at the level
of $\sim30$\,m\AA\ in at least some Pleiades stars on $\sim 10$ year timescales.

The evidence for variation in the \ki\ line is also difficult to
find. There is some indication that the S93a EWs are systematically
lower than those presented here, which make the results for Hz 1124
intriguing. It seems that the EW was larger in the S93a observation, 
perhaps behaving in a similar fashion to the \lii\ EW. The other six
stars in common, would be quite comparable if I have generally
overestimated the continuum level around the \ki\ line by $\sim 3\%$
with respect to S93a. Differences of this order are just plausible
because of the many telluric features close to the line, the different
resolutions of the observations and perhaps an under-estimation of
scattered light in S93a's echelle observations.

Lastly, we can look at chromospheric activity. Soderblom et al. (1993b)
list a number of
measurements for some stars taken in 1988 and 1990, 
and our additional measurements paint a
consistent picture of low-level variability of less than a factor of
two on timescales of 1-10 years. Interestingly, Hz 1124 is the most
constant star in our sample (in terms of percentage variation) 
in contrast to the results from the other lines.

\section{Discussion}

The new measurements of \lii\ EW presented here were obtained with
identical instrumentation and taking care to avoid any systematic
errors. The conclusion drawn from a small sample is that the \lii\ EW
is not variable on one year timescales at more than a level of about
10-15\,m\AA. For a late G/early K star with an effective temperature
of 5000\,K the curves of growth presented in S93a can be used to estimate
that this would produce a spread in Li {\em abundance} of $\leq 0.1$
dex for the most Li abundant stars in the Pleiades and perhaps $\leq 0.15$
dex for the least Li abundant Pleiades stars at this temperature. 
Variations at this level cannot therefore explain the vast majority of
the more than one order of magnitude apparent Li abundance scatter in the
Pleiades late G/early K stars.

On longer timescales of $\sim 10$ years, the data do support a limited
amount of \lii\ variability, perhaps as much as an rms of 30\,m\AA.
While this is still far too small (0.2-0.3 dex) to explain the Li abundance scatter,
it does agree with the kind of \lii\ variations which have been seen on
some, but not all, young, nearby, rapidly rotating K-type Pleiades analogues. For
example Hussain et al. (1997) measure a 45\,m\AA\ EW variation with
rotational phase in AB
Dor, and similar variations were recorded by
Jeffries et al. (1994) for BD$+22^{\circ}$~4409. These variations were
interpreted in terms of rotational modulation caused by enhanced \lii\ from
asymmetrically distributed starspots.  It is tempting to speculate that
the variability seen here is caused by the same phenomenon. However, it is then
difficult to understand (other than by appealing to the small sample
size) why no variation is seen on yearly timescales. It has been
clearly demonstrated (for example by Barnes et al. 1998) that spot
patterns on these young stars may only remain coherent for a month or
less and in any case, it is unlikely that the stars were viewed at the same
rotational phase (rotation periods are too uncertain to check this point).
A more likely explanation could lie in magnetic activity
cycles. A cycle length of order 11 years (like the Sun) 
would mean that that any effects of
plage or spot activity spread over the whole stellar surface may not
show great variation from year to year, but possibly would over the
course of many years or decades. There is evidence that some, but not
all, cool stars share solar-like magnetic activity cycles (Baliunas et
al. 1998). Few stars as active as the Pleiades G/K stars have been
studied in such detail, although AB Dor's overall spot coverage does reveal changes
on $\sim 10$ year timescales (Anders, Coates \& Thompson 1992).

A more sceptical interpretation is that much of the long term variation
is caused by uncertainties in the setting of continuum levels prior to
EW measurement. Even continuum level errors at the 1-2 percent level could result in 
measured EW errors of about 10-20\,m\AA. Effects of this size are
certainly plausible given the S/N and different
spectral resolutions of the datasets. Some support for this explanation
is offered by the correlation between \cai\ and \lii\ EW discrepancies
in some, but not all, cases. It is therefore still possible that
some of the variation is genuine.

The results presented here also suggest that chromospheric activity
variations on timescales of 1-10 years are limited to 20-30 percent in
these active, young stars.  This result is not entirely surprising,
although some rare exceptions were found by Soderblom et al. (1993b) in
their more extensive H$\alpha$ survey of the Pleiades. A similar lack
of long term variability has been found in the coronal emission of
Pleiades stars at X-ray wavelengths (Gagn\'{e}, Caillault \& Stauffer
1995), in contrast to the order of magnitude variability in X-ray
activity exhibited by the Sun over the course of its activity
cycle. The lack of chromospheric activity variability, combined with
the order of magnitude dispersion in the chromospheric activity of the
Pleiades stars considered here (presumably due to different rotation
rates), make it difficult to argue against active regions distributed
over the entire surface being responsible for at least a proportion of
the \lii\ EW scatter. Stuik et al. (1997) have modelled the \lii\ and
\ki\ sensitivity to homogeneously distributed spots and plages on the
surfaces of stars similar to those considered here. They find that
\lii\ {\em and} \ki\ EW scatters of $\sim 100$\,m\AA\ are possible
given the appropriate combinations of spot and plage filling factors.

Stuik et al. (1997) have also shown that
the \ki\ line is an extremely good proxy for the \lii\ line and mimics
its response to different atmospheric stratifications. S93a noticed that
there was a spread in \ki\ EW at a given $B-V$ and this point has
been stressed by Carlsson et al. (1994) and Stuik et al. (1997) as a
reason to be sceptical about whether the spread in \lii\ EW genuinely
reflects different Li abundances. 
The new data presented here
confirm this result and also that those stars with the largest \lii\ and
\ki\ EWs tend to be the most chromospherically active and rapidly
rotating. Indeed, S93a did not present \ki\ measurements for Hz 174 and Hz
2407 which are among the most rapidly rotating stars of their spectral
type in the Pleiades, and have the highest \ki\ EWs in this
temperature range. These data reinforce the conclusion that there is a
factor of two spread in the \ki\ EW at a given $B-V$ (in the range
$0.8<B-V<0.95$), compared with a factor of three spread in the \lii\
EW. While this \ki\ EW scatter remains unexplained, it is
dangerous to attribute the spread in \lii\ EWs to abundance
differences. Note though, that at cooler temperatures the spread in
\lii\ EW grows while the \ki\ EW spread remains roughly constant, but the
correlation between rotation and \lii\ EW also becomes much weaker.
Perhaps this indicates, in agreement with the predictions of current
models (Pinsonneault et al. 1998), that during the PMS phase at least,
rotationally dependent non-standard mixing is {\em not} the architect
of spreads in the \lii\ EW.

\section{Conclusions}

I have observed a small sample of Pleiades late G and early K stars
with consistent instrumentation and analysis techniques to look for
variation in the strength of the \lii\ 6708\AA\ line on one year
timescales. Variability on ten year timescales has also been searched
for using previously published data. At the same time I have measured
the strength of chromospheric activity and the \ki\ 7699\AA\ line,
which might plausibly have shown correlated variations. 
The following
conclusions can be drawn.
\begin{enumerate}
\item
The detection
of \lii\ EW variability comparable to the spread in \lii\ EW in this
temperature range would have immediately rejected the hypothesis that
the spread reflected a genuine Li abundance scatter. I detect no
variability on one year timescales that could correspond to
to more than about 0.1
dex in the deduced Li abundances.
\item
On 10 year timescales there is some evidence for \lii\ EW variability
which could produce 0.2-0.3 dex Li abundance spreads. Because these
data were taken with different instruments and spectral resolutions, I
remain sceptical about whether such variations are real, due to the
difficulty in applying consistent continuum placements.
\item
The first two conclusions lead us to believe that whatever causes the
scatter in \lii\ EWs, it is not something that varies significantly on
1-10 year timescales.
\item
There is only 20-30 percent variability in the levels of chromospheric
activity in this sample, despite a dispersion in chromospheric activity of
an order of magnitude. This makes it difficult to rule out
uniformly distributed magnetically active regions as a cause of \lii\ line
strength scatter.
\item
The data reinforce the conclusion that the strengths of {\em both} the
\lii\ 6708\AA\ and \ki\ 7699\AA\ lines increase with rotation rate and
chromospheric activity at a given colour. It seems unsafe to conclude that Li
depletion in the effective temperature range 4800-5200\,K 
is correlated with rotation until the large scatter in the \ki\
EWs has been satisfactorily explained.
\end{enumerate}

A number of useful extensions to this work are
apparent. Ideally, the sample size should be increased and extended to
include cooler Pleiads. It would also be most advantageous to repeat the
observations presented here in about 5 years time, using the same
instrumentation if possible to avoid systematic errors. Simultaneous
photometry or doppler imaging could be used to determine the
actual surface coverage of spots or active regions at the time of
observation.

\section*{Acknowledgements}
The Isaac Newton Telescope is operated on the island of La Palma by the
Isaac Newton Group in the Spanish Observatorio del Roque de los
Muchachos of the Instituto de Astrofisica de Canarias.  RDJ
acknowledges the travel and subsistence support of the UK Particle
Physics and Astronomy Research Council (PPARC) and the assistance of
A. Ford in obtaining the November 1998 dataset.  Computational work was
performed on the Keele node of the PPARC funded Starlink network.

\nocite{gagne95}
\nocite{anders92pasa}
\nocite{baliunas98}
\nocite{houdebine95}
\nocite{giampapa84}
\nocite{basri91}
\nocite{odell93}
\nocite{robinson86}
\nocite{eibe99}
\nocite{patterer93}
\nocite{fekel96}
\nocite{pallavicini93}
\nocite{russell96}
\nocite{barnes98}
\nocite{hussain97}
\nocite{pinsonneaultli98}
\nocite{randichaperli98}
\nocite{butler87}
\nocite{duncan83}
\nocite{boesgaard91}
\nocite{ryan95}
\nocite{soderblom93pleiadesli}
\nocite{soderblom93rot}
\nocite{chaboyer95}
\nocite{martin96}
\nocite{balachandran96}
\nocite{deliyannis97}
\nocite{carlsson94}
\nocite{martin98}
\nocite{ventura98}
\nocite{jeffriesblanco199}
\nocite{jeffries94}
\nocite{bonifacio97}
\nocite{stuik97}
\nocite{deliyannis90}
\nocite{pinsonneault90}
\nocite{mendes98}
\nocite{pinsonneault92}
\nocite{schatzman93}
\nocite{montalban96}

\bibliographystyle{mn}
\bibliography{iau_journals,master}

\label{lastpage}
\end{document}